\documentclass[acmsmall]{acmart}
\AtBeginDocument{%
  }

\setcopyright{acmlicensed}
\copyrightyear{2025}
\acmYear{2025}
\acmDOI{XXXXXXX.XXXXXXX}

\usepackage{algorithm}
\usepackage{algorithmic}
\usepackage{booktabs}
\usepackage{graphicx}
\begin{document}

\title{A Constrained Multi-Agent Reinforcement Learning Approach to Autonomous Traffic Signal Control}


\author{Anirudh Satheesh}
\affiliation{%
  \institution{University of Maryland - College Park}
  \city{College Park}
  \country{USA}}
\email{anirudhs@terpmail.umd.edu}

\author{Keenan Powell}
\affiliation{%
  \institution{University of Maryland - College Park}
  \city{College Park}
  \country{USA}}
\email{kpowell1@terpmail.umd.edu}

\renewcommand{\shortauthors}{Satheesh \& Powell}

\begin{abstract}
  Traffic congestion in modern cities is exacerbated by the limitations of traditional fixed-time traffic signal systems, which fail to adapt to dynamic traffic patterns. Adaptive Traffic Signal Control (ATSC) algorithms have emerged as a solution by dynamically adjusting signal timing based on real-time traffic conditions. However, the main limitation of such methods is they are not transferable to environments under real-world constraints, such as balancing efficiency, minimizing collisions, and ensuring fairness across intersections. In this paper, we view the ATSC problem as a constrained multi-agent reinforcement learning (MARL) problem and propose a novel algorithm named Multi-Agent Proximal Policy Optimization with Lagrange Cost Estimator (MAPPO-LCE) to produce effective traffic signal control policies. Our approach integrates the Lagrange multipliers method to balance rewards and constraints, with a cost estimator for stable adjustment. We also introduce three constraints on the traffic network: GreenTime, GreenSkip, and PhaseSkip, which penalize traffic policies that do not conform to real-world scenarios.  Our experimental results on three real-world datasets demonstrate that MAPPO-LCE outperforms three baseline MARL algorithms by across all environments and traffic constraints (improving on MAPPO by $12.60\%$, IPPO by $10.29\%$, and QTRAN by $13.10\%$). Our results show that constrained MARL is a valuable tool for traffic planners to deploy scalable and efficient ATSC methods in real-world traffic networks. We provide code at \url{https://github.com/Asatheesh6561/MAPPO-LCE}.
\end{abstract}





\maketitle

\section{Introduction}
\label{sec: introduction}
Traditional traffic signal systems, which operate on pre-programmed, fixed schedules, are often inadequate in addressing the dynamic nature of urban traffic flow due to an inability to adapt to constantly changing traffic patterns. This can result in longer waiting times and unfair traffic distributions across intersections \citep{atsc}. To combat the limitations of traditional fixed-time traffic signal systems, Adaptive Traffic Signal Control (ATSC) methods have been developed to adjust signal timing based on real-time traffic conditions dynamically. However, while ATSC methods hold promise in reducing congestion in busy intersections, there are still uncertainties about their deployment in real-world environments. One challenge is balancing efficiency while minimizing vehicle collisions and other hazards \citep{ESSA2020105713}. Another challenge is maximizing the fairness of each intersection, or ensuring that the green times (amount of time the current traffic light is green) for different lanes are the same on average \citep{10.1109/ITSC48978.2021.9564847}. In general, these challenges highlight the ongoing struggles with incorporating constraints into ATSC methods that accurately reflect the demands of real-world environments.

Previous works on ATSC use the observations of the intersections to form traffic control policies, such as SOTL \citep{Cools_2007}. However, these are heuristic-based and cannot adapt to more complex traffic environments. Additionally, they do not consider how current actions can affect future states, which hinders long-term outcomes. Reinforcement Learning (RL) has also been used to develop autonomous traffic control methods by optimizing over current and future states \citep{10.1145/3357384.3357900}. This includes actor-critic methods \citep{ASLANI2017732} and policy gradient methods \citep{mousavi2017trafficlightcontrolusing, 8832406} on single intersections \citep{NEURIPS2020_29e48b79} and multi-intersection environments \citep{10.1145/3357384.3357902, Chen_Wei_Xu_Zheng_Yang_Xiong_Xu_Li_2020}. RL has also been used for non-traditional intersections such as roundabouts \citep{8917519} and dynamical lane changing systems \citep{Zhou_2022}.

Due to the exponentially growing action space of reinforcement learning as the number of intersections increases, it becomes difficult to learn effective single-agent RL policies that can adapt to non-stationary environments like traffic signal control. As such, some works formulate ATSC as a decentralized Multi-Agent Reinforcement Learning (MARL) problem, using several agents to represent each intersection instead of one agent as a global traffic controller. This allows each intersection to act as its own local RL agent under partial observability and maximize its utility along with the global utility \citep{Zhou_2022, 9295660, wang2023novelmultiagentdeeprl}. Additional work serves to improve baseline MARL algorithms by improving sample efficiency \citep{huang2023fairnessaware}, or adding information to the state space to mitigate partial observability, such as communication methods \citep{DBLP:conf/ijcai/JiangQS0Z22} and environment modeling \citep{10.1007/978-3-031-70381-2_11, 10.1145/3292500.3330949}.

Due to the efficacy of MARL in solving high-dimensional traffic control problems and current struggles with incorporating constraints that reflect real-world environments, we propose a constrained MARL algorithm named Multi-Agent Proximal Policy Optimization with Lagrange Cost Estimator (MAPPO-LCE). Specifically, the algorithm uses the Lagrange multipliers method to balance the constraints with maximizing rewards and a cost estimator function to update the Lagrange multiplier. 

Our contributions can be summarized as follows:
\begin{enumerate}
    \item We define three constraint functions: GreenSkip, GreenTime, and PhaseSkip, which penalize policies that do not reflect real-world scenarios.
    \item We propose a constrained MARL algorithm for multi-intersection traffic control. 
    \item We show experimentally that MAPPO-LCE outperforms three baseline MARL algorithms on three different datasets.
    \item Our results show that constrained MARL can be a valuable tool for traffic planners to deploy ATSC methods in real-world traffic networks to reduce congestion.
\end{enumerate}

\section{Related Work}
\label{sec: related work}
In this section, we discuss recent work on MARL algorithms and general constraints for ATSC.
\subsection{MARL for ATSC}
\label{MARL for ATSC}
Recent work uses multi-agent reinforcement learning to model traffic signal control, with each agent controlling one intersection under partial observability. \citet{9295660} developed independent and joint Advantage Actor-Critic (A2C) algorithms for ATSC with a centralized critic in a distributed setting. \citet{9484451} also leverages A2C in a multi-agent setting, using decentralized critics for each agent in a distributed network. In addition to on-policy algorithms, previous works use multi-agent off-policy algorithms for ATSC. For example, \citet{10422534} uses Nash Q-Learning to alleviate the large state-action space from traditional MARL algorithms. \citet{wang2023novelmultiagentdeeprl} improves on this by using a Deep Q-Network \citep{mnih2013playingatarideepreinforcement} with Friend Q-Learning \citep{10.5555/645530.655661} to achieve better coordination between agents.

Other ways to improve MARL algorithms in ATSC are to include additional information in each agent's observation space to create more informed policies. However, including more information does not always lead to better results, as this can require more parameters and a slower convergence rate \citep{zheng2019diagnosingreinforcementlearningtraffic}. Thus, selecting the right information to include between agents is crucial for performance. \citet{huang2023fairnessaware} use a model-based approach by learning a global probabilistic dynamics model along with the policy, which generates a prediction of the next states as additional information. This method is purely decentralized, where there is no interaction between agents. Thus, \citet{DBLP:conf/ijcai/JiangQS0Z22} develops UniComm, a method that computes only the necessary information between neighbor agents, which is used in their UniLight algorithm to calculate Q values for each agent.
\subsection{Constraints for ATSC}
\label{Constraints for ATSC}
Solving environments with incorporated constraints is difficult due to balancing rewards and costs from the constraints. Constrained Reinforcement Learning (CRL) is an active research area in RL that solves such environments by developing algorithms that exclusively learn policies that are both effective and satisfy the constraints (e.g. safety, fairness, etc.) \citep{gu2022multiagentconstrainedpolicyoptimisation, achiam2017constrainedpolicyoptimization, lu2021decentralized}. \citet{achiam2017constrainedpolicyoptimization} develops a Constrained Policy Optimization (CPO) algorithm to learn policies under constraints, and \citet{gu2022multiagentconstrainedpolicyoptimisation} expands this into a multi-agent setting with MACPO and MAPPO-Lagrange. \citet{tabas2023interpreting} improve upon MACPO by developing a primal-dual optimization framework and parameterizing each agent with a neural network.
 
In ATSC, there is minimal work on incorporating constraints into the environment to develop policies closer to real-world scenarios. \citet{10.1109/TITS.2024.3352446} partitions the traffic network topology to alleviate scalability issues with MARL, but this only constrains the state space, not the action space. \citet{10.1145/3676169} use the CRL framework with the amount of emissions as the constraint and develop a Soft Actor-Critic algorithm to balance rewards with constraints. However, this is a single-agent setting, which poses scalability issues as the number of intersections increases. \citet{10422440} models traffic environment constraints in a multi-agent setting, but this work models agents as the vehicles around one intersection, instead of each intersection being an agent.
Finally, \citet{10.1109/ITSC48978.2021.9564847} creates two fairness constraints for the ATSC problem, one delay-based metric which is meant to diminish the number of vehicles experiencing significantly longer waiting times and another throughput-based metric which attempts to give equal weighting to all traffic flows by extending concepts from computer networking. However, this is also a single agent setting in a more simplistic environment and is focused more specifically on fairness between the North-South and East-West traffic flows instead of constraints under general traffic network topologies. 
\section{Preliminaries}
\label{sec: Preliminaries}
In this section, we define the Constrained Markov Game, the RL environment, and our constraints for ATSC. 

\subsection{Constrained Markov Game for ATSC}
\label{subsec: Constrained Markov Game for ATSC}
We can model ATSC as a constrained Markov Game 
\citep{qu2024safetyconstrainedmultiagentreinforcement, wang2024safe} which can be represented by the tuple $M = \langle\mathcal{N}, S, \{O_i\}_{i \in \mathcal{N}}, \{A_i\}_{i \in \mathcal{N}}, \mathcal{T}, r, \Omega, C, c, \gamma \rangle$, where $\mathcal{N} = \{1, 2, . . . , n\}$ is a set of $n$ agents; $S$ is the state space; $O = \times_{i \in \mathcal{N}}O_i$ is the joint observation space, where $O_i$ is the observation space of agent $i$; $A = \times_{i \in \mathcal{N}}A_i$ is the joint action space, where $A_i$ is the action space of agent $i$; $T : S\times A\times S \rightarrow [0, 1]$ is probabilistic state transition function; $R$ is the reward function; $\Omega : S \times A \times O \rightarrow [0, 1]$ is space of conditional observation probabilities \(\left(\Omega(s', a, o) = P(o | s', a)\right)\); $C : S \times A \rightarrow \mathbb{R}$ is the cost function; and $c$ is the cost limit. Since this is a decentralized Markov Game, the reward function for each agent is the same, e.g. \(R = R_i \, \forall i \in \mathcal{N}\). MARL algorithms for constrained Markov Games aim to search for policy $\pi$ that solves this constrained optimization problem:
\begin{align*}
    &\max_\pi\,\;  \mathbb{E}_{\left(s_t \sim S, a_t \sim \pi \right)}\left[\sum_{t=0}^{\infty}\gamma^t r(s_t, a_t)\right], \\
    &\text{s.t.}\,\;\;\;\;\mathbb{E}_{\left(s_t \sim S, a_t \sim \pi \right)}\left[\sum_{t=0}^{\infty} \gamma^t C(s_t, a_t)\right] < c
\end{align*}
In the ATSC problem specifically, the elements of the environment are defined as:
\begin{itemize}
    \item Agents: Each agent is responsible for controlling traffic lights at one intersection.
    \item Observation: The observation of each agent is composed of the characteristics of the corresponding intersection. Specifically, each intersection has 12 road links (vehicles turning left, right, and going straight in each cardinal direction), and each road link contains the number of vehicles moving, the number of vehicles waiting, the traffic light phase, and the number of vehicles in each lane, as well as the speed and location of each vehicle in the lane. 
    \item Actions: As shown in Figure 1 and Figure 2, there are eight phases that describe combinations of traffic lights that can be green simultaneously. At each timestep, the intersection can choose one of these phases as an action.
    \item STATE: The state is the combination of all observations at the current time step.
    \item STATE Transition: After an action is selected at each time step, vehicles are allowed to move if the corresponding traffic light is green for a short period \(T_g\). While the environment does not directly represent yellow lights, before changing phases, all lights that would be turned on/off are turned to red for a brief period \(T_y\) before the lights of the new phase are turned to green.
    \item Reward: Each agent will receive a global reward \(\lambda_fR_f + \lambda_wR_w\), where \(R_f\) is the total number of vehicles moving, \(R_w\) is the total number of vehicles waiting, and \(\lambda_f\) and \(\lambda_w\) are hyperparameters. 
\end{itemize}
For more information on environment parameters, refer to Appendix A.
\subsection{Environment Constraints}
\label{subsec: Environment Constraints}
We develop three environment constraints on each intersection that reflect real-world environments named GreenTime, PhaseSkip, and GreenSkip \citep{FGSV_RiLSA_2010}. These constraints also help to promote fair treatment of all vehicles by the agents by reducing differences in waiting times between directions and encouraging agents to take all possible actions.
\begin{itemize}
    \item GreenTime: Each light \(l\) should be green for no more than $G_{max\; time}$ before turning red to prevent long waiting times from other lanes, and model light cycles in the real world. Each time step that a light is on increases its GreenTime value by 1, and when it is turned off its GreenTime value is set to 0. Right-turn lights are ignored for this constraint, as they are always treated as being on. 
    \begin{align}
        G_{time}(l) \leq G_{max\;time}
    \end{align}
    \item PhaseSkip: The state of each traffic light follows one of a specific, pre-determined set of phases (see Figure 2). No phase should be skipped consecutively more than $P_{max\;skip}$ times. Each time the phase changes, the new phase has its PhaseSkip value set to 0, and all phases other than the new phase and the old phase have their PhaseSkip values incremented by 1. This is a way of somewhat closely approximating how traffic cycles work in the real world, as well as being an indirect way of promoting the agent to give equal attention to all lanes.
    \begin{align}
        P_{skips}(p) \leq P_{max\;skips}
    \end{align}
    \item GreenSkip: Similar to the phase constraint, no individual light \(l\) should be skipped consecutively more than $G_{max\;skips}$ times. Each time the phase changes, each light turned on in the new phase has its GreenSkip value set to 0, and all lights not on in the new phase or the old phase have their GreenSkip values incremented by 1. This is a direct way of promoting fairness by reducing the variance in waiting times among all lanes. Right-turn lights are also ignored for this constraint.
    \begin{align}
        G_{skips}(l) \leq G_{max\;skips}
    \end{align}
\end{itemize}
Each agent is constrained according to Eqns 1-3. The penalty associated with each constraint is the average across all lights:
\begin{align}
    \frac{\sum_{i \in \mathcal{N}}\sum_{l}\frac{1_c}{n_l(i)}}{|\mathcal{N}|}
\end{align}
where \(1_c\) is an indicator function that checks whether the constraint is satisfied, $i$ is the intersection, $|\mathcal{N}|$ is the number of agents, $l$ is a specific light at the intersection the agent controls, and $n_l(i)$ is the total number of lights at the intersection that particular agent controls. Note that for the PhaseSkip constraint, we sum over the phases and divide by the total number of phases. For our experiments, the number of lights is always 12 and the number of phases is always equal to 8, as all the intersections in our environment have 4 roads of 3 lanes with 8 distinct phases. The exact algorithms for calculating each constraint can be referenced in Algorithms \ref{alg: greentime}, \ref{alg: phaseskip}, and \ref{alg: greenskip}. 

\begin{algorithm}
    \caption{GreenTime Calculation}\label{alg: greentime}
    \begin{algorithmic}[1]
        \FOR{\textbf{time} = $1$ to $N$}
            \FOR{\textbf{light} \textbf{in} lights}
                \IF{\textbf{light} is ON in the current phase}
                    \STATE \textbf{green\_time[light]} $\gets$ green\_time[light] + 1
                \ELSE
                    \STATE \textbf{green\_time[light]} $\gets 0$
                \ENDIF
            \ENDFOR
        \ENDFOR
    \end{algorithmic}
\end{algorithm}

\begin{algorithm}
    \caption{PhaseSkip Calculation}\label{alg: phaseskip}
    \begin{algorithmic}[1]
        \FOR{\textbf{time} = $1$ to $N$}
            \IF{\textbf{new\_phase} $\neq$ \textbf{old\_phase}}
                \FOR{\textbf{phase} \textbf{in} phases}
                    \IF{(\textbf{phase} $\neq$ old\_phase) and (\textbf{phase} $\neq$ new\_phase)}
                        \STATE \textbf{phase\_skips[phase]} $\gets$ phase\_skips[phase] + 1
                    \ENDIF
                \ENDFOR
                \STATE \textbf{phase\_skips[new\_phase]} $\gets 0$
            \ENDIF
        \ENDFOR
    \end{algorithmic}
\end{algorithm}

\begin{algorithm}
    \caption{GreenSkip Calculation}\label{alg: greenskip}
    \begin{algorithmic}[1]
        \FOR{\textbf{time} = $1$ to $N$}
            \IF{\textbf{new\_phase} $\neq$ \textbf{old\_phase}}
                \FOR{\textbf{light} \textbf{in} lights}
                    \IF{(\textbf{light} is RED in old\_phase) and (RED in new\_phase)}
                        \STATE \textbf{green\_skips[light]} $\gets$ green\_skips[light] + 1
                    \ELSE
                        \STATE \textbf{green\_skips[light]} $\gets 0$
                    \ENDIF
                \ENDFOR
            \ENDIF
        \ENDFOR
    \end{algorithmic}
\end{algorithm}

\section{Method}
\label{sec: Method}
In this section, we describe our constrained multi-agent reinforcement learning algorithm: Multi-Agent Proximal Policy Optimization with Lagrange Cost Estimator (MAPPO-LCE).
\subsection{Multi-Agent Proximal Policy Optimization with Lagrange Cost Estimator}
\label{subsec: MAPPO-LCE}
Constrained optimization problems are typically of the form 
\begin{align*}
    &\max_x f(x) \\
    & \text{s.t. } g(x) \leq c
\end{align*}
which can be solved by the Lagrange multiplier method 
\begin{align}
    \mathcal{L}(x; \lambda) = f(x) - \lambda(g(x) - c)
\end{align}
where \(\mathcal{L}(x; \lambda)\) is a new optimization objective to maximize and \(\lambda > 0\) is the Lagrange multiplier.
\citep{Bertsekas_1996}. Thus, for the constrained MARL problem,
\begin{align}
    &\max_{\pi_{\theta}}\,\;  \mathbb{E}_{\left(s_t \sim S, a_t \sim \pi_{\theta} \right)}\left[\sum_{t=0}^{\infty}\gamma^t r(s_t, a_t)\right], \\
    &\text{s.t.}\,\;\;\;\;\mathbb{E}_{\left(s_t \sim S, a_t \sim \pi_{\theta} \right)}\left[\sum_{t=0}^{\infty} \gamma^t C(s_t, a_t)\right] < c
\end{align}
we can formulate it with a Lagrangian where 
\begin{align}
    &f(x) = \mathbb{E}_{\left(s_t \sim S, a_t \sim \pi \right)}\left[\sum_{t=0}^{\infty}\gamma^t r(s_t, a_t)\right] \\
    &g(x) = \mathbb{E}_{\left(s_t \sim S, a_t \sim \pi \right)}\left[\sum_{t=0}^{\infty} \gamma^t C(s_t, a_t)\right]
\end{align}
In MAPPO-LCE, we use a reward critic and a cost critic, \(V^r_{\phi_r}\) and \(V^c_{\phi_c}\), for estimating the discounted cumulative reward and discounted cumulative cost, respectively. We choose to build off of MAPPO because we require only one actor and one critic model during training and inference, which reduces the computation and memory requirements of the algorithm. Instead of training on every step, we also collect a dataset \(D\) containing rollout data every episode: \(\{s_t, r_t, c_t, s_{t+1}\}\). After \(B\) episodes, we update the policy. Similar to MAPPO-Lagrange \citep{gu2022multiagentconstrainedpolicyoptimisation}, we aim to minimize the following loss:
\begin{align}
    \mathcal{L}(\pi_{\theta}) = \mathcal{L}_{r}(\pi_{\theta}) - \lambda\mathcal{L}_c(\pi_{\theta})
    \label{eqn: actor loss}
\end{align}
where \(\mathcal{L}_r\) and \(\mathcal{L}_c\) are the MAPPO \citep{yu2022surprising} actor losses with an unclipped critic loss term:
\begin{align}
\mathcal{L}_r(\pi_{\theta}) = 
& \mathbb{E}_{s_t \sim D, a_t \sim \pi_{\theta}} \bigg[ \min \big( \rho_t A^r_t, \text{clip}(\rho_t, 1 \pm \epsilon) A^r_t \big) \bigg] \notag \\
&+ \beta \mathbb{E}_{s_t \sim D} \bigg[\frac{1}{2} \| V^r_{\phi_r}(s_t) - r_t \|^2 \bigg]\\
\mathcal{L}_c(\pi_{\theta}) = 
&\mathbb{E}_{s_t \sim D, a_t \sim \pi_{\theta}} \bigg[ \min \big( \rho_t A^c_t, \text{clip}(\rho_t, 1 \pm \epsilon) A^c_t \big) \bigg] \notag \\
&+ \beta\mathbb{E}_{s_t \sim D} \bigg[ \frac{1}{2} \| V^c_{\phi_c}(s_t) - c_t \|^2 \bigg]
\end{align}
In these formulations, \(\rho_t\) is the importance sampling ratio 
\[
    \rho_t = \frac{\pi_{\theta}(a_t|s_t)}{\pi_{\theta_{\text{old}}}(a_t|s_t)}
\]
\(A^c_t\) and \(A^r_t\) are the cost advantage and reward advantage functions respectively, and \(\epsilon\) is the clipping parameter. Here, we abuse notation and say that \(A^c_t = A^c_t(s_t, a_t)\) and \(A^r_t = A^r_t(s_t, a_t)\). We also update each critic by the temporal difference error (TDE):
\begin{align}
    &\mathcal{L}_{\phi_r} = \mathbb{E}_{(s_t,\, s_{t+1}) \sim 
 D}\left[r_t + \gamma V^r_{\phi_r}(s_{t+1}) - V^r_{\phi_r}(s_{t})\right] 
 \label{eqn: reward critic loss}\\
    &\mathcal{L}_{\phi_c} = \mathbb{E}_{(s_t,\, s_{t+1}) \sim 
 D}\left[c_t + \gamma V^c_{\phi_c}(s_{t+1}) - V^c_{\phi_c}(s_{t})\right]
 \label{eqn: cost critic loss}
\end{align}
In MAPPO-Lagrange \citep{gu2022multiagentconstrainedpolicyoptimisation}, the Lagrange multiplier \(\lambda\) is updated by the mean of the cost advantage function \(A^c_t\). This works in theory because the cost advantage function measures how much constraint violation occurs in a certain state when taking a particular action, compared to the mean constraint violation over all actions. Thus, if taking an action in a state results in a negative cost advantage, \(\lambda\) should be increased to alleviate this. However, since the advantage function only converges during the policy learning process, it may take many iterations to accurately estimate the constraint violation. During this time, the estimates can be unstable and potentially incorrect. To address this issue, we incorporate a Lagrange Cost Estimator, inspired by \citet{qu2024safetyconstrainedmultiagentreinforcement} to provide more stable and reliable estimates of constraint violations. This cost estimator quickly learns the cost dynamics within the first few iterations to accurately predict the cost, and then updates \(\lambda\). We train the cost estimator \(\theta_C\) by minimizing the following loss:
\begin{align}
    \mathcal{L}_{\theta_C} = \|\theta_C(s_t, a_t) - c_t\|^2,\, s_t \sim D, a_t \sim \pi_\theta
    \label{eqn: cost estimator}
\end{align}
Finally, we update \(\lambda\) with the following loss to ensure that the constraint function is satisfied under the cost limit \(c\):
\begin{align}
    \mathcal{L}_{\lambda} = \mathbb{E}_{s_t \sim D, a_t \sim \pi_\theta} \left[-\lambda(\theta_C(s_t, a_t) - c)\right]
    \label{eqn: lambda update}
\end{align}
as the loss is minimized when the estimated cost is much less than the cost limit. 
One consideration is that instead of updating the policy in a fully online manner, we perform rollouts of the MARL policy for one episode and store the trajectories (containing the state, action, reward, cost, and next state) in a replay buffer. Then during training, we randomly sample from this replay buffer. The main advantage is that each agent learns from a more diverse set of environment updates and reduces the variance of gradient updates. Additionally, we clamp \(\lambda\) to be greater than zero to ensure that the policy is always penalized when the constraints are violated. Finally, to allow for smoother transitions between updates in the actor model and the critic models, we perform soft updates using the frozen versions of the models used in the Temporal Difference Error calculations. We display the full algorithm in Algorithm \ref{alg: MAPPO-LCE Algorithm}.

\begin{algorithm}
\caption{MAPPO-LCE Algorithm}
\begin{algorithmic}
\STATE Initialize replay buffer $\mathcal{D}$, policy parameters $\theta$, critic networks $V_\phi^r$, $V_\phi^c$, cost network $\theta_C$, and Lagrange multiplier $\lambda$.
\FOR{each episode} 
    \FOR{each time step $t$} 
        \STATE Select action $a_t$ = $\pi_\theta(s_t)$
        \STATE Execute joint action $a_t$ at state $s_t$
        \STATE Observe reward $r_t$, cost $c_t$, and next state $s_{t+1}$
        \STATE $\mathcal{D} \leftarrow \mathcal{D} \cup (s_t, a_t, r_t, c_t, s_{t+1})$
    \ENDFOR
\STATE Sample batch $\mathcal{B}$ from $\mathcal{D}$
\STATE $\theta \leftarrow \theta - \alpha \nabla\mathcal{L}(\pi_\theta)$ by Equation \ref{eqn: actor loss}
\STATE $\phi^r \leftarrow \phi^r - \alpha \nabla \mathcal{L}_{\phi_r}$ by Equation \ref{eqn: reward critic loss} 
\STATE $\phi^c \leftarrow \phi^c - \alpha \nabla \mathcal{L}_{\phi_c}$ by Equation \ref{eqn: cost critic loss}
\STATE $\theta_C \leftarrow \theta_C - \alpha_{\theta_C}\nabla \mathcal{L}_{\theta_C}$ by Equation \ref{eqn: cost estimator}
\STATE $\lambda \leftarrow \lambda - \alpha_{\lambda}\nabla_{\lambda} \mathcal{L}_{\lambda}$ by Equation \ref{eqn: lambda update}
\STATE Clamp to ensure $\lambda \geq 0$ 
\STATE Soft update actor and critic parameters:
\[
\theta \leftarrow \tau \theta + (1 - \tau) \theta', \quad 
\phi_r \leftarrow \tau \phi_r + (1 - \tau) \phi_r', \quad \phi_c \leftarrow \tau \phi_c + (1 - \tau) \phi_c'
\]
\ENDFOR
\end{algorithmic}
\label{alg: MAPPO-LCE Algorithm}
\end{algorithm}

\section{Experiments}
\label{sec: Experiments}
In this section, we outline our experimental details, including the environment configurations and explanation of baseline algorithms.
\subsection{Environment Setup}
\label{subsec: Environment Setup}
We run our experiments on MAPPO-LCE and related baselines on the CityFlow environment \citep{zhang2019cityflow}, which is a scalable and realistic traffic simulator due to its C++ backend. Additionally, it is compatible with several multi-agent RL algorithms by integrating with the Gymnasium library \citep{brockman2016openai}. From \citet{10.1145/3357384.3357902}, there are three publicly available datasets collected from real-world traffic data from Hangzhou, China (HZ); Jinan, China (JN); and New York, USA (NY). Details of each environment are located in \ref{tab: environment stats}.

To evaluate the performance of each of the MARL algorithms, we use three evaluation metrics:
\begin{itemize}
    \item Test Reward: The test reward is the same as the training reward: \(\lambda_fR_f + \lambda_wR_w\).
    \item Average Delay: The average delay is the average delay across all vehicles, which is the total travel time minus the expected travel time for each vehicle. The expected travel time is the estimated time the vehicle should finish its route if there were no traffic lights.
    \item Throughput: The throughput of the environment is the number of vehicles that complete their routes before the episode ends.
\end{itemize}

\begin{table}[h]
\centering
\begin{tabular}{lccc}
\toprule
 & HZ & JN & NY\\
 \cmidrule(rl){2-4}
Number of Intersections       & 16     & 12     & 48     \\

Number of Lanes               & 3      & 3      & 3      \\

Total Number of Vehicles      & 2983   &  6295  & 2824  \\

Time Steps (s)       & 3600   & 3600   & 3600   \\
\bottomrule
\end{tabular}
\caption{Summary of Traffic Metrics for HZ, JN, and NY.}
\label{tab: environment stats}
\end{table}

\subsection{Baseline Methods}
\label{subsec: Baseline Methods}
In this work, we compare our algorithm to three baseline MARL algorithms: Independent Proximal Policy Optimization (IPPO) \citep{de2020independent}, Multi-Agent Proximal Policy Optimization (MAPPO) \citep{yu2022surprising}, and QTRAN \citep{son2019qtran}. This set of algorithms allows us to test both on-policy algorithms (IPPO, MAPPO), and off-policy algorithms (QTRAN).

\begin{itemize}
    \item IPPO \citep{de2020independent}: IPPO treats each agent as its independent local RL agent to maximize local rewards. This transforms the problem into \(|\mathcal{N}|\) independent single-agent PPO rollouts. \\
    \item MAPPO \citep{yu2022surprising} MAPPO joins the actions of each agent into a single joint action vector, and each agent shares an actor network and a critic network to update the policy. \\
    \item QTRAN \citep{son2019qtran}: QTRAN develops an unstructured value function factorization, which allows for more generalizable decentralized execution of MARL problems.
\end{itemize}
All baseline algorithms were implemented or derived from the ePYMARL library \citep{papoudakis2021benchmarking}. For each algorithm, the total reward at time step \(t\) is \(r_t - \zeta c_t\), where \(r_t\) and \(c_t\) are the rewards and costs at time step \(t\), and \(\zeta\) is a hyperparameter that trades off maximizing the reward and satisfying the constraints. 
All experiments were conducted on a single RTX A5000 GPU.
\section{Results}
\label{sec: Results}
In this section, we show the results of our algorithm on several environment configurations and perform ablation studies to highlight the advantages of specific components of MAPPO-LCE.

\subsection{Main Results}
\label{subsec: Main Results}
\begin{figure*}[h]
\centering
\includegraphics[width=0.9\textwidth]{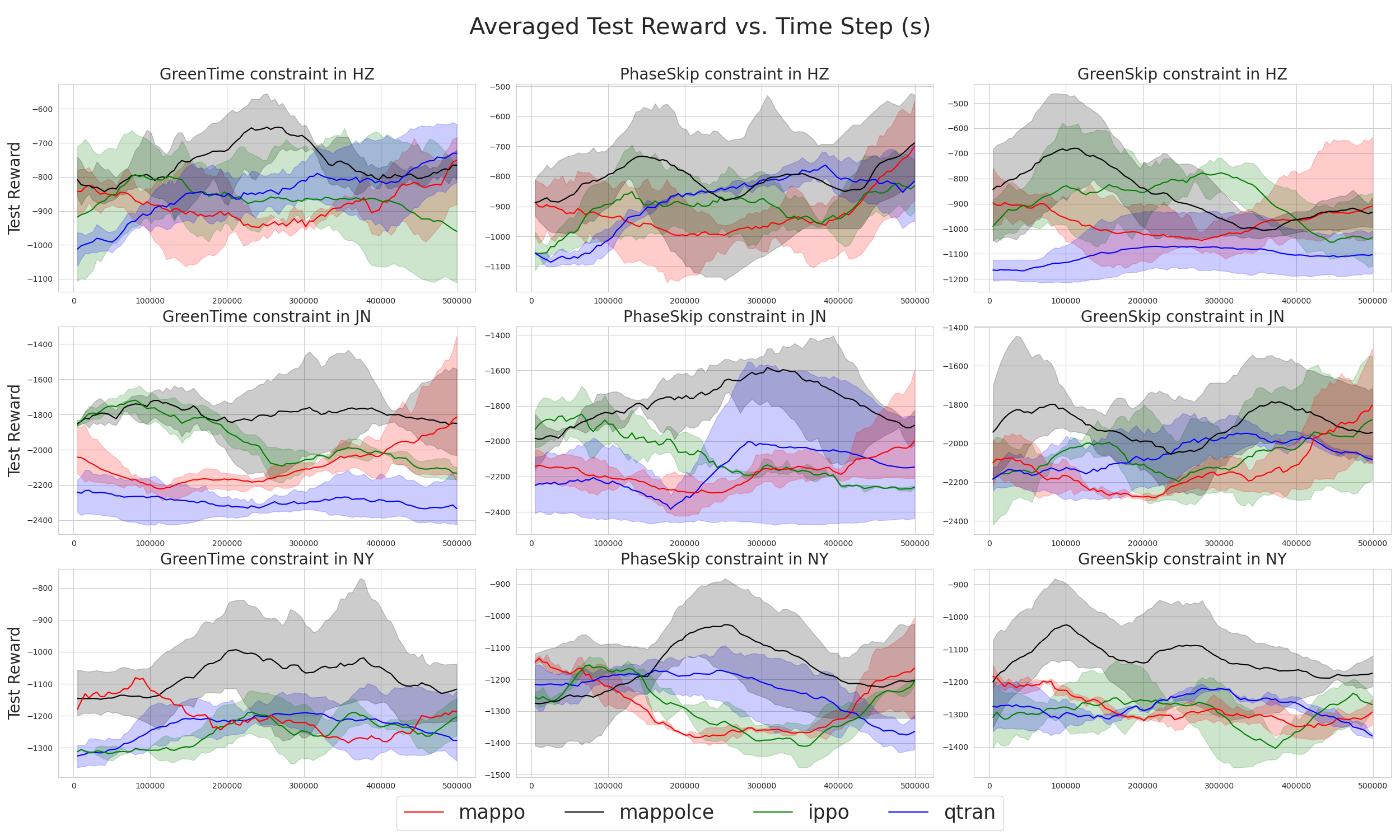}
\caption{Plot of test reward on the over 500,000 timesteps for the MAPPO-LCE algorithm compared to baseline algorithms across all environments and constraints.}
\label{fig: Test Reward Results}
\end{figure*}

\begin{table}[h]
\centering
\begin{tabular}{llp{3cm}p{3cm}p{3cm}}
\toprule
 Environment & Constraint & \multicolumn{3}{c}{MAPPO-LCE \% Reward increase over comparison algorithms } \\
 \cmidrule(rl){3-5}
 & & MAPPO & IPPO & QTRAN \\
 \midrule
 HZ & GreenTime & \textbf{13.86\%} & 12.15\% & 10.61\% \\
 HZ & PhaseSkip & \textbf{12.88\% }& 10.99\% & 8.74\% \\
 HZ & GreenSkip & 11.27\% & 2.55\% & \textbf{21.58\%} \\
 JN & GreenTime & 14.22\% & 7.65\% & \textbf{21.74\%} \\
 JN & PhaseSkip & \textbf{18.57\%} & 15.03\% & 17.95\% \\
 JN & GreenSkip & \textbf{10.69\%} & 7.73\% & 7.39\% \\
 NY & GreenTime & 10.4\% & \textbf{14.15\%} & 12.62\% \\
 NY & PhaseSkip & 9.46\% & \textbf{9.52\% }& 5.55\% \\
 NY & GreenSkip & 12.05\% & \textbf{12.83\%} & 11.75\% \\
\bottomrule
\end{tabular}
\caption{Comparison of the Test Reward metric between MAPPO-LCE and MARL baseline algorithms across all constraint and environment combinations.}
\label{tab: test reward stats}
\end{table}

\begin{table}[h]
\centering
\begin{tabular}{llp{3cm}p{3cm}p{3cm}}
\toprule
Environment & Constraint & \multicolumn{3}{c|}{MAPPO-LCE \% Throughput increase over comparison algorithms } \\
\cmidrule(rl){3-5}
 & & MAPPO & IPPO & QTRAN \\
 \midrule
 HZ & GreenTime & \textbf{11.4\%} & -5.93\% & -0.59\% \\
 HZ & PhaseSkip & \textbf{10.66\%} & -5.44\% & 1.81\% \\
 HZ & GreenSkip & 11.88\% & -13.73\% & \textbf{34.89\%} \\
 JN & GreenTime & 23.9\% & 10.45\% & \textbf{57.99\%} \\
 JN & PhaseSkip & 29.77\% & 11.3\% & \textbf{29.86\%} \\
 JN & GreenSkip & \textbf{15.42\%} & 0.46\% & 11.85\% \\
 NY & GreenTime & 40.25\% & 7.43\% & \textbf{50.88\%} \\
 NY & PhaseSkip & \textbf{78.22\%} & 18.86\% & 44.02\% \\
 NY & GreenSkip & \textbf{74.15\%} & 15.6\% & 63.91\% \\
\bottomrule
\end{tabular}
\caption{Comparison of the Test Throughput metric between MAPPO-LCE and MARL baseline algorithms across all constraint and environment combinations. Here, a higher metric indicates a better policy.}
\label{tab: test throughput stats}
\end{table}

\begin{table}[h]
\centering
\begin{tabular}{llp{3cm}p{3cm}p{3cm}}
\toprule  
Environment & Constraint & \multicolumn{3}{c}{MAPPO-LCE \% Average Delay decrease} \\ 
\cmidrule(rl){3-5}
 & & MAPPO & IPPO & QTRAN \\ 
 \midrule
 HZ & GreenTime & 24.23\% & 8.56\% & \textbf{20.44\%} \\
 HZ & PhaseSkip &\textbf{ 13.73\%} & -2.97\% & 13.31\% \\
 HZ & GreenSkip & 10.83\% & -17.89\% & \textbf{24.72\%} \\
 JN & GreenTime & 15.64\% & 8.19\% & \textbf{24.44\%} \\
 JN & PhaseSkip & 19.27\% & 10.52\% & \textbf{19.78\%} \\
 JN & GreenSkip & \textbf{10.6\%} & 2.11\% & 8.51\% \\
 NY & GreenTime & 8.36\% & 3.94\% & \textbf{10.47\%} \\
 NY & PhaseSkip & \textbf{13.41\%} & 6.71\% & 11.52\% \\
 NY & GreenSkip & \textbf{11.65\%} & 4.79\% & 11.94\% \\ \bottomrule
\end{tabular}
\caption{Comparison of the Test Average Delay metric between MAPPO-LCE and MARL baseline algorithms across all constraint and environment combinations. Note that a higher metric here is better due to an average delay decrease.}
\label{tab: test average delay stats}
\end{table}

The results of the algorithms on the three environments are shown in Figure \ref{fig: Test Reward Results} and Table \ref{tab: test reward stats}. In these figures, we include the results of each algorithm on the test reward function defined in Section \ref{subsec: Environment Setup}.

As shown in Table \ref{tab: test reward stats}, MAPPO-LCE outperforms all three comparison algorithms in every combination of environment and constraint that was tested on. While some of the other algorithms come close to the performance of MAPPO-LCE on specific setups (e.g. IPPO on HZ GreenSkip or QTRAN on NY PhaseSkip), taking the average across all runs yields a 12.60\% improvement over MAPPO, a 10.29\% improvement over IPPO, and a 13.10\% improvement over QTRAN. Additionally, taking the average over the different constraints, MAPPO-LCE sees a 13.05\% improvement with GreenTime, a 12.08\% improvement with PhaseSkip, and a 10.87\% improvement with GreenSkip. The slight decay in improvement with PhaseSkip or GreenSkip is likely due to them over-constraining the action space and too strongly encouraging the model to switch into unoptimal phases too often, but even with those restrictions, the model still sees consistent improvements.

One additional notable result from Figure \ref{fig: Test Reward Results} is that our algorithm more consistently outperforms the comparison algorithms as the complexity of the environment increases (we consider the NY environment the most complex, as it contains 3-4 times more agents than the HZ and JN environments). This is because in harder environments with more agents, policies are more heavily penalized for violating constraints. MAPPO-LCE's ability to adaptively handle constraints through a learnable Lagrange multipliers parameter enables it to more effectively balance performance while adhering to constraints across all agents. This underscores the scalability of MAPPO-LCE, which is essential for incorporating MARL policies in real-world settings. 

A further analysis of the results can be done by looking at Tables \ref{tab: test throughput stats} and \ref{tab: test average delay stats}, which show other important metrics for traffic systems (Figures \ref{fig: Test Throughput Results} and \ref{fig: Test Average Delay Results}). MAPPO-LCE is sometimes outperformed in either or both of these metrics by other algorithms - however, this only occurs in the HZ environment, which is the simplest. MAPPO-LCE still outperforms all three comparison algorithms in general, however, it has much smaller margins of improvement over IPPO. In terms of Throughput, it achieves a 32.85\% improvement over MAPPO, a 32.73\% improvement over QTRAN, but only a 4.33\% improvement over IPPO. In terms of Average Delay, it achieves a 14.19\% improvement over MAPPO, a 16.13\% improvement over QTRAN, but only a 2.66\% improvement over IPPO. This is likely because the traffic environment is largely independent, as while nearby traffic lights exhibit some dependencies due to their close proximity, the influence of distant traffic lights is minimal and delayed. Thus, an independent policy algorithm like IPPO can effectively capture localized decision-making dynamics while avoiding unnecessary coordination. Additionally, \citet{de2020independent} shows that IPPO's performance on fully cooperative tasks can exceed other on-policy algorithms like MAPPO and Q-function factorization methods such as QMIX \citep{rashid2018qmixmonotonicvaluefunction} and QTRAN due to the importance of policy value clipping. However, IPPO still lags behind MAPPO-LCE due to the Lagrange Cost Estimator's ability to guide the policy in the implicitly constrained state and action space.

We also show the averaged individual constraint values for each of the algorithms in Figures \ref{fig: green time constraint values}, \ref{fig: green skip constraint values}, and \ref{fig: phase skip constraint values}, where each constraint value of a particular constraint ranges from 0-1 representing the proportion of lights/phases in violation of the constraint, with 0 being no violations and 1 being all possible violations. In almost all environment and constraint combinations, the value of the constraint violation is lowest for MAPPO-LCE compared to the other baseline algorithms. Additionally, QTRAN and MAPPO's performance decrease is supported by the high constraint violations, especially for GreenSkip and PhaseSkip, as the number of constraint violations increases significantly during training. However, while IPPO has much better performance in each environment than QTRAN and MAPPO (but not as much as MAPPO-LCE), it still has large constraint violations, especially concerning the GreenSkip constraint. Finally, we note that MAPPO-LCE's improved performance on the harder JN and NY datasets is supported by the lower constraint values for those two datasets, which continues to demonstrate the algorithm's scalability. 

\subsection{Further Results}
\begin{figure*}[h]
\centering
\includegraphics[width=0.9\textwidth]{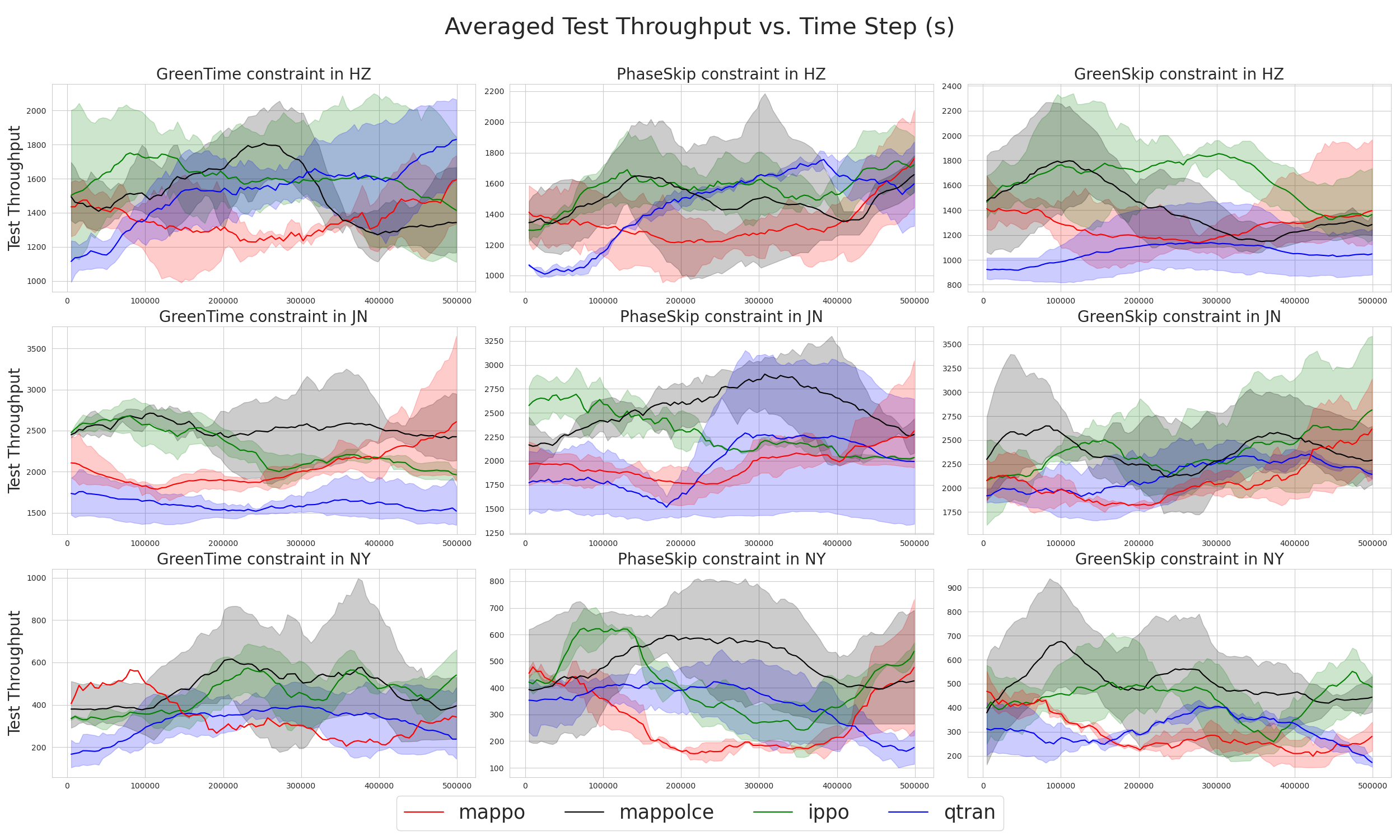}
\caption{Plot of throughput on the test set over 500,000 timesteps for the MAPPO-LCE algorithm compared to baseline algorithms across all environments and constraints.}
\label{fig: Test Throughput Results}
\end{figure*}

\begin{figure*}[h]
\centering
\includegraphics[width=0.9\textwidth]{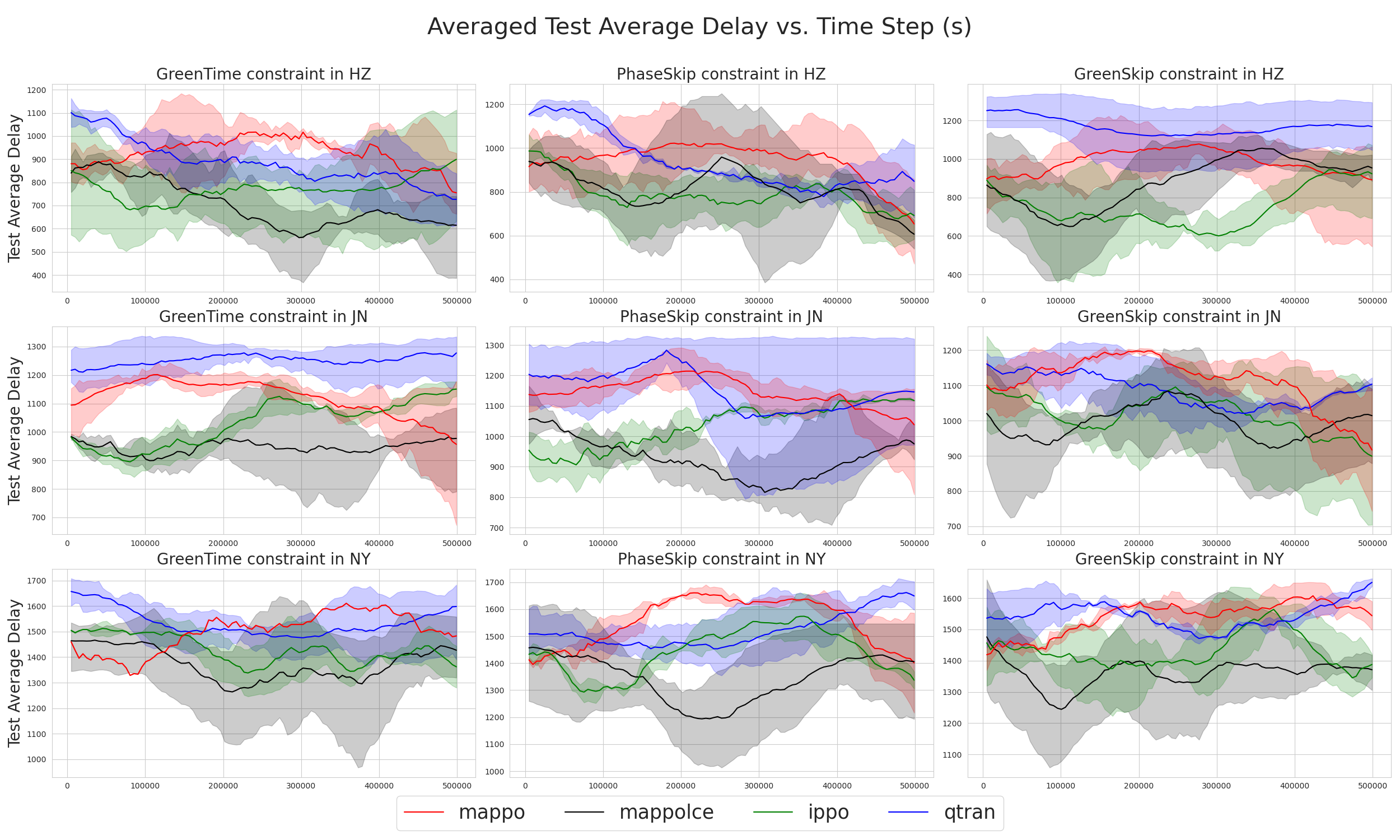}
\caption{Plot of average delay on the test set over 500,000 timesteps for the MAPPO-LCE algorithm compared to baseline algorithms across all environments and constraints.}
\label{fig: Test Average Delay Results}
\end{figure*}

\begin{figure}[h]
\centering
\includegraphics[width=0.9\textwidth]{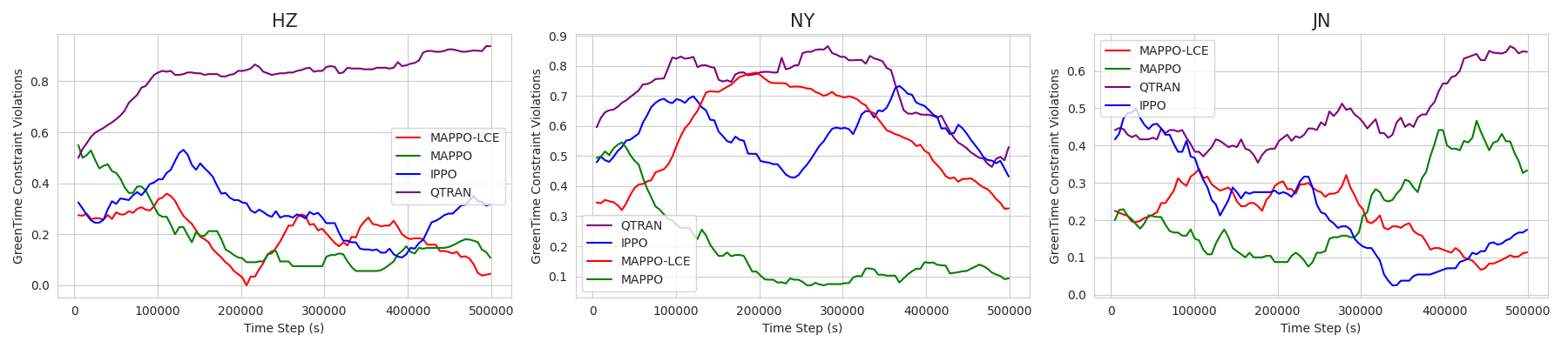}
\caption{Plot of Green Time constraint values over all environments and algorithms.}
\label{fig: green time constraint values}
\end{figure}

\begin{figure}[h]
\centering
\includegraphics[width=0.9\textwidth]{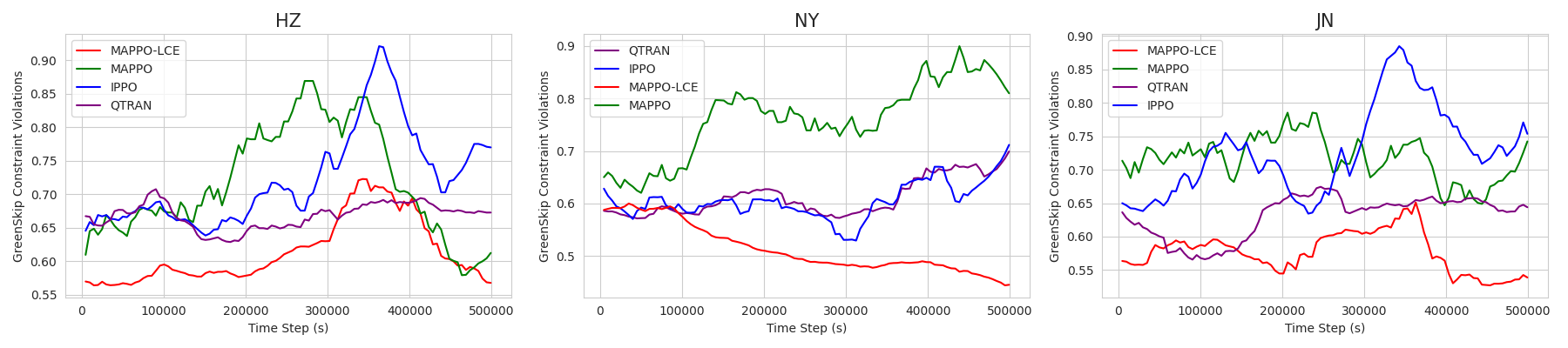}
\caption{Plot of Green Skip constraint values over all environments and algorithms.}
\label{fig: green skip constraint values}
\end{figure}

\begin{figure}[h]
\centering
\includegraphics[width=0.9\textwidth]{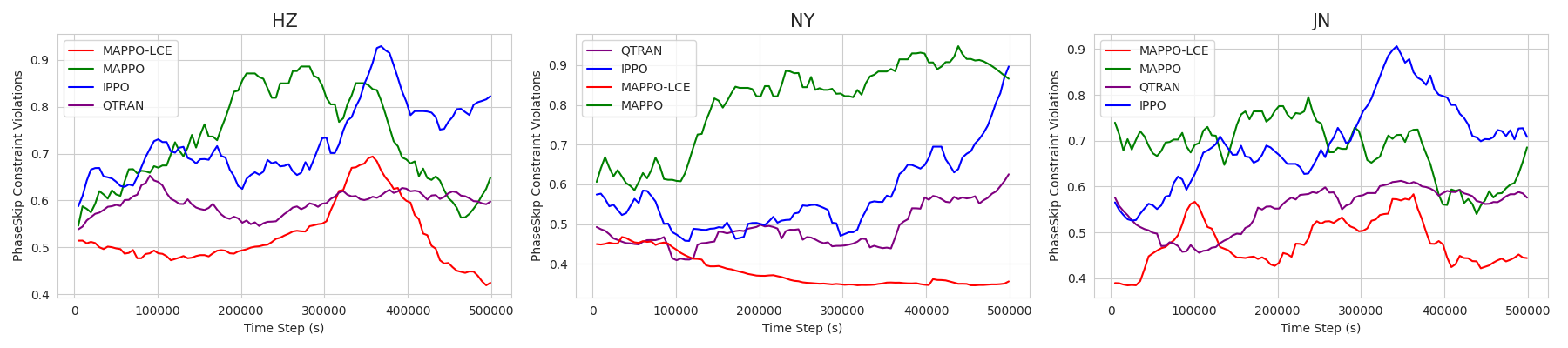}
\caption{Plot of Phase Skip constraint values over all environments and algorithms.}
\label{fig: phase skip constraint values}
\end{figure}

\begin{figure}[h]
\centering
\includegraphics[width=0.9\textwidth]{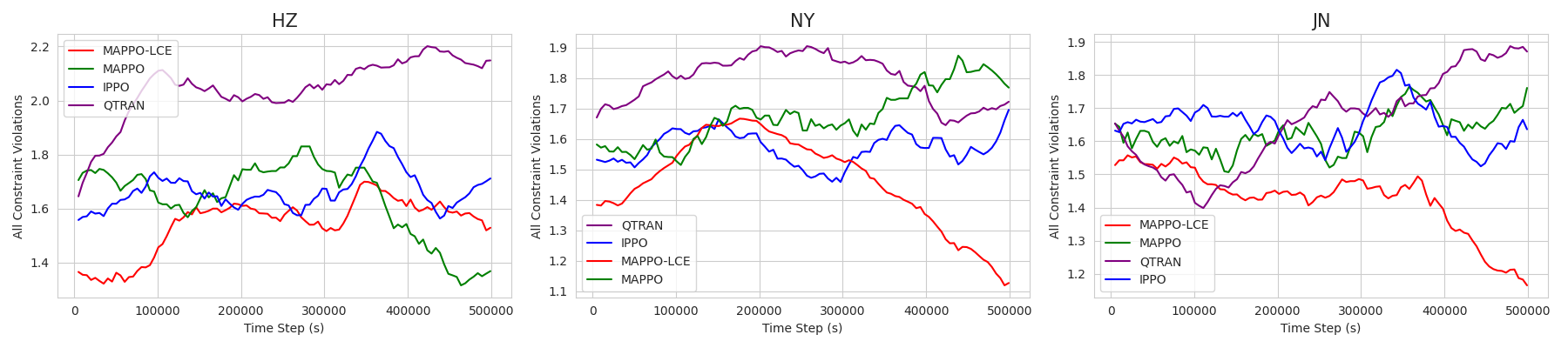}
\caption{Plot of summed constraint values over all environments and algorithms.}
\label{fig: all constraint values}
\end{figure}

In Figures \ref{fig: Test Throughput Results} and \ref{fig: Test Average Delay Results}, we show the throughput and average delay results on the test set across timesteps for the combinations of environments and constraints. Additionally, we provide experiment results for the setting where all constraints are combined linearly to test each algorithm's ability to handle multiple constraints in Tables \ref{tab: test all reward stats}, \ref{tab: test all throughput stats}, and \ref{tab: test all delay stats}. When evaluating this challenging setting where all constraints are combined, MAPPO-LCE continues to achieve the highest overall reward, outperforming MAPPO, IPPO, and QTRAN by 6.43\%, 0.22\%, and 4.13\%, respectively (see Tables \ref{tab: test all reward stats}, \ref{tab: test all throughput stats}, and \ref{tab: test all delay stats}). While the performance margins are smaller than in individual constraint settings, this is expected due to the inherent complexity of optimizing conflicting objectives (GreenTime, PhaseSkip, and GreenSkip simultaneously). Still, even in this more difficult scenario, MAPPO-LCE remains the most robust algorithm, demonstrating its ability to learn balanced policies under multiple constraints. 
Interestingly, IPPO occasionally achieves marginally better throughput (2.25\%) and delay (0.58\%) than MAPPO-LCE in this setting, likely due to the increased optimization instability introduced by competing constraints. However, IPPO’s gains come at the cost of higher constraint violations (see Figure \ref{fig: all constraint values}). Figure \ref{fig: all constraint values} shows the summed constraint values across the test runs on each environment. Notice that while MAPPO-LCE only attains a slightly lower amount of constraint violations compared to IPPO on HZ (with IPPO trending lower at the end), it by far trends the lowest on both of the more complex JN and NY datasets. This again highlights MAPPO-LCE's ability to reliably enforce constraints without sacrificing quality reward optimization, as it maintains competitive or better metrics on reward, throughput, and delay, while much more strongly minimizing constraint violations.

\begin{table}[h]
\centering
\begin{tabular}{lp{3cm}p{3cm}p{3cm}}
\toprule
 Environment & \multicolumn{3}{c}{MAPPO-LCE \% Test Reward increase over comparison algorithms} \\
 \cmidrule(rl){2-4}
 & MAPPO & IPPO & QTRAN \\
 \midrule
 HZ & \textbf{10.31\%} & 0.63\% & 6.76\% \\
 JN & \textbf{3.5\%} & 1.0\% & 0.1\% \\
 NY & 5.48\% & -0.96\% & \textbf{5.53\%} \\
\bottomrule
\end{tabular}
\caption{Comparison of the Test Reward between MAPPO-LCE and MARL baseline algorithms using a sum of all constraints for each traffic environment.}
\label{tab: test all reward stats}
\end{table}

\begin{table}[h]
\centering
\begin{tabular}{lp{3cm}p{3cm}p{3cm}}
\toprule
 Environment & \multicolumn{3}{c}{MAPPO-LCE \% Throughput increase over comparison algorithms} \\
 \cmidrule(rl){2-4}
 & MAPPO & IPPO & QTRAN \\
 \midrule
 HZ & \textbf{17.75\%} & -0.73\% & 0.37\% \\
 JN &\textbf{ 7.25\% }& 2.66\% & 1.39\% \\
 NY & \textbf{36.93\%} & -8.69\% & 24.84\% \\
\hline
\end{tabular}
\caption{Comparison of the Test Throughput metric between MAPPO-LCE and MARL baseline algorithms using a sum of all constraints for each traffic environment.}
\label{tab: test all throughput stats}
\end{table}

\begin{table}[h]
\centering
\begin{tabular}{lp{3cm}p{3cm}p{3cm}}
\toprule
 Environment & \multicolumn{3}{c}{MAPPO-LCE \% Average Delay decrease over comparison algorithms} \\
 \cmidrule(rl){2-4}
 & MAPPO & IPPO & QTRAN \\
 \midrule
 HZ & \textbf{14.87\%} & -0.99\% & 6.12\% \\
 JN & \textbf{4.74\%} & 1.41\% & 1.44\% \\
 NY & 5.19\% & -2.17\% & \textbf{5.23\%} \\
\bottomrule
\end{tabular}
\caption{Comparison of the Test Average Delay metric between MAPPO-LCE and MARL baseline algorithms using a sum of all constraints on each traffic environment.}
\label{tab: test all delay stats}
\end{table}

\section{Future Work}
For future work, one idea is to incorporate communication between agents, as such communication is already implemented in practice via connected vehicle technology \citep{s21227712}. This would allow traffic lights to communicate information such as the number of vehicles traveling from one intersection to another, the average travel time, and other vital data that could help discover more optimal policies. This also can be easily represented with an underlying Graph Neural Network (GNN), which means we can add message loss to the overall loss as both the GNN and the neural networks in each algorithm use backpropagation. Additionally, following from Section \ref{subsec: Main Results}, excessive communication may be unnecessary and potentially prohibitive. Thus, leveraging pruning strategies on the communication graph, such as \citep{das2020tarmactargetedmultiagentcommunication, niu2021multi, Ding_Du_Ding_Guo_Zhang_2024, hu2024learningmultiagentcommunicationgraph} would help avoid redundant communication that would perversely affect agent learning. 

Another idea is to expand our constraints. Formulating our current constraints as hard constraints and adding additional soft constraints such as variance in throughput or waiting times could more closely represent real-world environments while creating a model that values fairness and safety. In addition, we could add further constraints by expanding the environment to include different types of vehicles, such as buses, ambulances, and trams, to develop more generalizable traffic management strategies that accommodate diverse transportation needs. Adding more significant constraints on their delay and waiting time could lead to more robust constraints that better reflect real-life scenarios. 

Expanding the ways our algorithm treats constraints and is able to incorporate them is also another route for future expansion. For example, one aspect of our algorithm is that the constraint value is globalized and applied to all agents equally, which means we only need one cost critic and cost estimator. However, this means we cannot model each agent to correct any individual constraint violations. Future work serves to accurately incorporate such fine-grained control over constraint violations without the significant overhead of training cost estimators for each agent.

While ATSC is a partially observable Markov Game, a final idea is to give each agent a better idea of their surroundings through expectation alignment. ELIGN \citep{ma2022elignexpectationalignmentmultiagent} is a method for multi-agent expectation alignment that aligns the shared expectations of an agent to its actual actions through an intrinsic reward. Predicting neighboring agents' actions in a second-order theory of mind approach allows for better coordination and can easily be added to existing methods to find more optimal policies. In the ATSC problem, this may allow each agent to predict swells or dips in traffic before they reach the intersection that the agent controls, further increasing its ability to make realistic traffic policies. 

\section{Conclusion}
In this paper, we focus on finding scalable algorithms for the Adaptive Traffic Signal Control problem in real-world traffic environments. We propose a novel algorithm, MAPPO-LCE, for constrained multi-agent reinforcement learning. We expand upon Multi-Agent Proximal Policy Optimization (MAPPO) by incorporating elements of MAPPO-Lagrangian \citep{gu2022multiagentconstrainedpolicyoptimisation} and introducing a Lagrange Cost Estimator to accurately predict constraints even under unstable conditions. While we only focused on three constraints, MAPPO-LCE can be used with any number of general traffic constraints and can be extended to any constrained MARL problem. Our experimental results using the CityFlow environment in multiple real-world settings show that MAPPO-LCE outperforms other baseline methods with suitable constraints. Our findings indicate that constrained multi-agent reinforcement learning can identify more optimal traffic policies for ATSC in real-world conditions and holds strong potential for real-world deployment.

% \begin{acks}

% \end{acks}

\bibliographystyle{ACM-Reference-Format}
\bibliography{submission}

\appendix
\section{Hyperparameter Selection}
\subsection{Algorithm Hyperparameters} 
We use the same hyperparameters since all of our baseline algorithms are adapted from the ePYMARL library \citep{papoudakis2021benchmarking}. We provide a full list of algorithmic environment hyperparameters in Table \ref{tab: hyperparameters}. For the constraint trade-off hyperparameter \(\zeta\), we set it to 0.2 for all constraints. We model the cost estimator as a Multi-Layer Perceptron (MLP), with two hidden layers and a hidden layer size of 128. We also use an Adam optimizer with a learning rate of \(10^{-4}\) to train the cost estimator. We set the cost limit for all experiments to \(0\).

\subsection{Environment Hyperparameters}
Each time step in the environment is composed of \(T_g\) inner steps to update the environment, which we set to 10. Thus, simulating for 500,000 steps is the same as simulating approximately 1400 episodes. After the policy selects an action, each inner step simulates 1 second of the environment. To simulate yellow lights without actually implementing them directly, each traffic light instead turns off all lights that would be switched between phases for $T_y$ time before fully turning them red or green. We set $T_y$ to 5 time steps, which is equivalent to 5 seconds. Each constraint also has a hyperparameter that controls its severity. For constraint thresholds, we set $G_{max\;time}$ to 40, $P_{max\;skips}$ to 16, and $G_{max\;skips}$ to 4.

\begin{table}[h]
\centering
\caption{Hyperparameter Comparison of RL Algorithms}
\label{tab:hyperparams}
\begin{tabular}{lcccc}
\hline
\textbf{Hyperparameter} & \textbf{IPPO} & \textbf{MAPPO} & \textbf{QTRAN} & \textbf{MAPPO-LCE} \\ \hline
Epsilon Start & 1.0 & 1.0 & 1.0 & 1.0 \\
Epsilon Finish & 0.05 & 0.05 & 0.05 & 0.05 \\
Epsilon Anneal Time & 500000 & 500000 & 500000 & 500000 \\
Learning Rate (lr) & 0.00005 & 0.0005 & 0.0005 & 0.00005 \\
Gamma & 0.985 & 0.985 & 0.985 & 0.985 \\
Hidden Dim & 128 & 128 & 32 & 128 \\
Grad Norm Clip & 10 & 10 & 5 & 10 \\
Critic Coef & 0.5 & 0.5 & - & 0.5 \\
Entropy Coef & 0 & 0 & - & 0 \\
Reg Coef & 0.01 & 0.01 & - & 0.01 \\
GAE Lambda & 0.95 & 0.95 & - & 0.95 \\
Mini Epochs & 2 & 2 & 1 & 2 \\
Eps Clip & 0.15 & 0.15 & - & 0.15 \\
Target Update Interval & 200 & 200 & 200 & 200 \\
Mixing Embed Dim & - & - & 64 & - \\
Opt Loss & - & - & 1 & - \\
Nopt Loss & - & - & 0.1 & - \\
Lambda Init & - & - & - & 0.01 \\
Lambda LR & - & - & - & 0.0001 \\ 
Batch Size & 8 & 8 & 8 & 8 \\
Buffer Size & 8 & 8 & 8 & 8 \\
\hline
\end{tabular}
\label{tab: hyperparameters}
\end{table}

\end{document}